\documentclass[useAMS,usenatbib]{mn2e}
\usepackage{epsfig}

\title[The Impact of a Supernova Explosion in a Very Massive Binary]{The Impact of a Supernova Explosion in a Very Massive Binary}

\author[J. Sato, M. Umemura and K. Sawada]{Jun'ichi Sato$^{1}$\thanks{Current address: Service d'Astrophysique, DSM/IRFU, UMR AIM, CEA-CNRS-Univ. Paris VII, Saclay, France;junichi.sato@ces.fr (JS)}, 
Masayuki Umemura$^{1}$ and Keisuke Sawada$^{2}$\\
$^{1}$Center for Computational Sciences, University of Tsukuba, Tsukuba 305-8577, Japan\\
$^{2}$Department of Aeronautics and Space Engineering, Tohoku University, Sendai 980-8579, Japan}

\begin{document}

\date{Accepted ........ Received ........ ; in original form ........}

\pagerange{\pageref{firstpage}--\pageref{lastpage}} \pubyear{2008}

\maketitle


\begin{abstract}
We consider the effect of a supernova (SN) explosion 
in a very massive binary that is expected to form in a portion
of Population III stars with the mass higher than 100$M_\odot$. 
In a Population III binary system, a more massive star can
result in the formation of a BH and a surrounding accretion disc.
Such BH accretion could be a significant source of the cosmic
reionization in the early universe. 
However, a less massive companion star evolves belatedly
and eventually undergoes a SN explosion, so that the accretion disc
around a BH might be blown off in a lifetime of companion star. 
In this paper, we explore the dynamical impact of a SN explosion on 
an accretion disc around a massive BH, and elucidate
whether the BH accretion disc is totally demolished or not. 
For the purpose, we perform three-dimensional hydrodynamic 
simulations of a very massive binary system, where we assume
a BH of $10^3 M_{\odot}$ that results from a direct collapse of a very
massive star and a companion star of $100 M_{\odot}$ that 
undergoes a SN explosion. 
We calculate the remaining mass of a BH accretion disc as a function
of time. As a result, it is found that a significant portion of gas disc
can survive through three-dimensional geometrical effects 
even after the SN explosion of a companion star. 
Even if the SN explosion energy is higher by two orders of magnitude
than the binding energy of gas disc, about a half of disc can be left over. 
The results imply that the Population III BH accretion disc 
can be a long-lived luminous source, and therefore
could be an important ionizing source in the early universe. 
\end{abstract}
\begin{keywords} 
accretion discs --- PopIII stars --- early universe  
--- hydrodynamics --- binaries: general --- cosmology: theory 
\end{keywords}


\section{Introduction}
\label{Sec1}
The studies on the formation of Population III (Pop III) stars
have shown that first stars in the universe are likely to 
be as massive as $100-1000M_\odot$ 
\citep{BCL99, Abel00,Abel02} or to
form in a bimodal initial mass function (IMF) with peaks of several $100M_\odot$ 
and $\sim 1M_\odot$ \citep{NU01}. 
Recently, it has been revealed that 
a significant fraction of Pop III stars can be expected 
to form in binary systems \citep{Saigo+04}. 
\citet{HW02} have studied the nucleosynthetic evolution of Pop III
stars, and have shown that Pop III stars in the mass range of 
$m \ga 260 M_\odot$
result in direct black hole formation, while the mass range of
$25 M_\odot \la m \la 140 M_\odot$ leads to the supernova (SN) explosions.
Hence, if there is mass difference in a Pop III binary, 
there can form a binary system composed of a BH with $m \ga 260 M_\odot$
and a massive star with $25 M_\odot \la m \la 140 M_\odot$.
Also, recent general relativistic simulations on a supermassive star 
($m \ga 10^3 M_\odot$)
have shown that if a star possesses 
large angular momentum, a massive gas disc is left over
around a forming black hole 
\citep{SS02, Shibata04, Shapiro04, SS04}. 
The mass fraction of gas disc is $10^{-4} - 10^{-1}$ 
of the initial stellar mass,
depending on the equation of state and the degree of rotation. 
This disc can fuel the BH in the later evolutionary stage.
Even if a disc left over is quite small, the Roche lobe overflow from
a companion star may also fuel the BH in a close binary system
\citep{Lawlor08}.

On the other hand, it is suggested that 
Pop III BH accretion may be an important clue 
for the reionization of the universe \citep{mad04,RO04b}. 
For a BH of $\approx 10^3 M_{\odot}$, 
the accretion disc can emit a blackbody radiation with 
an effective temperature of $10^{6-7} \mbox{ K}$,
if the mass accretion rate is close to the Eddigton rate 
(e.g. \citet{Kato+98}).
When the disc mass is several $10 M_{\odot}$,
the accretion timescale is several $10^{6} \mbox{ years}$. 
Therefore, the accreting BH can be an ultraviolet radiation source
in a longer timescale than the lifetime of a massive companion star. 
Then, the accreting BH can be a significant contributor 
for the early reionization\citep{HU07}. 
However, if the accretion disc is blown off by the SN explosion 
of a companion star, the accretion timescale may be too short 
to play a significant role for the reionization. 

In this paper, we explore the disruption and stripping of a gas disc
around a BH by a SN explosion of a massive companion star.
For the purpose, we perform three-dimensional hydrodynamic simulations
with a second-order scheme, AUSM-DV. In particular, to resolve the propagation
of a blast wave and the deformation of a gas disc with high accuracy,
three-dimensional generalized curvilinear coordinates are used. 
Such a three-dimensional simulation on the collision of SN blast wave 
with a BH accretion disc has been never performed so far. 
With the simulations, we study the stripping efficiency as a function of 
the kinetic energy of the SN blast wave, 
by changing the separation of binary and the mass of gas disc.
Then, the dependence on the ratio of the blast wave kinetic energy 
to the binding energy of gas disc is analysed in detail.

The paper is organized as follows. 
In Section 2, the details of present numerical model and 
method are presented. 
In Section 3, we show the numerical results for the deformation
and stripping of gas disc around a BH by the collision with
a SN blast wave. Also, the dependence on the model parameters
is studied, and the resultant remaining mass fraction of disc
is argued. Section 4 is devoted to conclusions.

\section{Numerical Model and Method}
\label{Sec2}

\subsection{Model}
\label{Sec2-1}

We consider a binary system composed of a black hole 
of $1000 M_{\odot}$ and a Pop III companion star of $100 M_{\odot}$. 
In the present analyses, the model parameters are the separation
between the BH and the companion star, 
the mass of gas disc surrounding the BH, and the density profile of gas disc. 
Based on the analysis by \citet{Saigo+04}, 
we set the binary separation, $a$, to be $700 \mbox{ AU}$ as a fiducial case, 
and a half or one-fifth of the fiducial separation is also examined. 
As for the mass of gas disc, considering the disc formation 
around a black hole \citep{SS02, Shibata04, Shapiro04, SS04}
and also the mass loss from a Pop III star before a SN explosion
\citep{HW02}, we set the disc mass to be $30 M_{\odot}$, 
$10 M_{\odot}$ or $3 M_{\odot}$.
Regarding the density distribution of gas disc, 
we assume a power-law type distribution like $\rho(r) \propto r^n$
on the equatorial plane. 
The power-law index $n$ is assumed to be $n=0$, $-1$, or $-2$. 
The temperature of gas disc is presumed to be $10^{4} \mbox{K}$, 
since the gas disc is irradiated by ultraviolet radiation from a companion
star before the explosion. 
The rotation of the disc is assumed to be Keplerian, because the self-gravity
is negligible in all the cases considered. 
In vertical directions of the disc, we determine the density distributions by
solving a hydrostatic balance. 
Also, taking the tidal effect into account, the gas disc is truncated at the 
tidal radius given by \cite{Paczynski77} and \cite{Boffin01}. 
The tidal radius is $\sim 0.7$ Roche lobe radius of the BH. 
For instance, the outer radius of the gas disc is set to be $280 \mbox{AU}$ 
for a fiducial case. 
The space out of the gas disc is filled with tenuous and high temperature gas,
since we cannot treat a vacuum in the present hydrodynamic scheme.
In order to minimize its dynamical effect, 
we set the density to be negligibly small.
Here, the density and temperature are assumed to be $2.5 \times 10^{-16} 
\mbox{g cm}^{-3}$ and $10^{7} \mbox{K}$, respectively.
The parameters adopted are summarized in Table \ref{tab1}. 

To simulate a SN explosion, we employ a method similar to
that taken by \citet{KY05}, and \citet{BYH03}, where the explosion energy
is inserted as the thermal energy of the gas ejected from a SN. 
The ejected gas has no initial velocity 
and is accelerated by the gas pressure. 
The temperature is assumed to be $1.0 \times 10^{8} \mbox{K}$ 
for all cases, and the density and extent of hot gas 
is adjusted so that the explosion energy is $10^{51} \mbox{erg}$ 
\citep{HW02}.
If the separation is shorter, 
the density of hot gas is higher and the extent is smaller.
For model A1, the density of the hot gas is set to be 
$\rho_{\rm SN}=1.0 \times 10^{-11} \mbox{g cm}^{-3}$ and
the extent is to be $50 \mbox{AU}$. 
We tested the cases of different density and extent 
and confirmed that the impact of a SN explosion on the gas disc 
does not have a strong dependence on the choices, but
is basically determined by the explosion energy.
Also, we regulate the duration of SN explosion 
so that the total mass ejected by SN explosion is
$50 M_{\odot}$ \citep{HW02}. 

\subsection{Numerical Method}
\label{Sec2-2}
We solve the three-dimensional Euler equation for inviscid gas.
Basic equations can be written in the conservative form as 
\begin{equation}
\frac{\partial \tilde{Q}}{\partial t}+
\frac{\partial \tilde{E}}{\partial x}+
\frac{\partial \tilde{F}}{\partial y}+
\frac{\partial \tilde{G}}{\partial z}+
\tilde{H}=0, 
\label{eq-1}
\end{equation}
where $\tilde{Q}$, $\tilde{E}$, $\tilde{F}$, $\tilde{G}$ and $\tilde{H}$ 
are respectively the following vectors, 
\begin{eqnarray}
&&
\tilde{Q}=
\left(
\begin{array}{c}
\rho \\
\rho u \\
\rho v \\
\rho w \\
e
\end{array}
\right)
,\quad
\tilde{E}=
\left(
\begin{array}{c}
\rho u \\
\rho u^{2}+p \\
\rho uv \\
\rho uw \\
(e+p)u
\end{array}
\right)
,\quad
\tilde{F}=
\left(
\begin{array}{c}
\rho v \\
\rho uv \\
\rho v^{2}+p \\
\rho vw \\
(e+p)v
\end{array}
\right)
,
\nonumber \\
&&
\tilde{G}=
\left(
\begin{array}{c}
\rho w \\
\rho uw \\
\rho vw \\
\rho w^{2}+p \\
(e+p)w
\end{array}
\right)
,\quad
\tilde{H}=
\left(
\begin{array}{c}
0 \\
\rho f_{x} \\
\rho f_{y} \\
\rho f_{z} \\
\rho \left( uf_{x}+vf_{y}+wf_{z} \right)
\end{array}
\right),
\label{eq-2}
\end{eqnarray}
(e.g, \citealt{Vinokur74}), 
where $\rho$ is the density of gas, $u$, $v$ and $w$ are 
respectively the $x$-, $y$- and $z$- components of velocities, 
and $f_{x}$, $f_{y}$ and $f_{z}$ are 
the $x$-, $y$- and $z$- components of gravity force by a black hole. 
As the gravity force, we take into account only the effect of black hole,
since it is a major component. 
The pressure $p$ and the total energy per unit volume $e$ 
are related by the equation of state, 
\begin{equation}
p=\left(\gamma-1\right)
\left\{e-\frac{\rho}{2}\left(u^{2}+v^{2}+w^{2}\right)\right\},
\label{eq-3}
\end{equation}
where $\gamma$ is the ratio of specific heats of gas, 
which is $\gamma=5/3$ here. 

The Reynolds number based on the molecular viscosity is very
large for the present case. Therefore, we have neglected the
effect of viscosity (and only considered the effect of ram
pressure), i.e. solved the Euler equations. However, we
should note that possible turbulence would provide a tendency
to reduce the remaining disc mass or to extinguish the disc.
Solving such case accounting for turbulent eddy viscosity
will be our future work
\citep{Takeda+85}. 

To solve the governing equation (\ref{eq-1}) numerically,
AUSM-DV scheme \citep{WL94} is employed. 
To properly adjust the boundary and solve the flow with high accuracy,
we use the three-dimensional generalized curvilinear coordinates as
shown in Fig. \ref{Fig1}. 
Because we assume the system is in a plane-symmetric with respect 
to the equatorial plane, we perform simulations in the computational domain 
that is composed of a hemisphere and an adjacent protuberant region.
The BH and the SN explosion are placed on the equatorial plane,
where the BH is located at the center of hemisphere, and 
the SN explosion occurs at the center of protuberant region. 
In Fig. \ref{Fig1}, only a half of the computational domain is 
shown after the domain is cut by a plane perpendicular to 
the equatorial plane that connects the BH and the SN explosion. 
The number of grid points is $51 \times 101 \times 31$. 
We treat a black hole as a small central hole with absorbing boundary, 
whose radius is $20 \mbox{AU}$ for the case of $a=700 \mbox{AU}$. 
The gas disc is set up around the small hole, and
the inner radius of the disc is assumed to be $50 \mbox{ AU}$. 
The grids are generated with higher resolution near the black hole 
and the equatorial plane of the gas disc. 
Such coordinates allow us to treat accurately the flow of the gas disc. 
The computations are carried out 
only in the upper half domain relative to the equatorial plane, 
because we assume that the system is symmetric about this plane. 
The leftward meshes are built up to represent a SN explosion. 
A free boundary condition on outer and inner boundaries is adopted
in these meshes. 

\section{Results}
\label{Sec3}

When a SN explosion produces a supersonic flow with the mass of 
$50 M_{\odot}$ and the velocity of $4000 \mbox{ km}/\mbox{sec}$, 
the kinetic energy of the blast wave is $\sim 1.1 \times 10^{50} \mbox{ erg}$
for the solid angle of a disc viewed from the SN explosion.
In the fiducial case, the binding energy of gas disc is evaluated 
to be $\sim 3.1 \times 10^{48} \mbox{ erg}$.
Hence, the kinetic energy of a SN blast wave is much larger than the disc binding 
energy. Then, the collision of the blast wave with a disc 
is anticipated to be devastating. 
However, the disc is deformed by the collision, and also streamlines are bent 
significantly. Therefore, it is unclear whether the whole of disc gas is 
stripped out by the collision of the SN blast wave.

\subsection{Disc Deformation and Stripping}
\label{deformation}

Figure \ref{Fig2} shows the numerical results at early stages for model A1.
The density distributions and velocity fields are shown in the $x-y$ plane 
with $z=0$ ({\it lower panels}) and in $x-z$ plane with
$y=0$ ({\it upper panels}). 
Figure \ref{Fig2}(a) shows the initial state, where 
high-density hot gas is ejected at $700 \mbox{AU}$ from a BH. 
The hot gas expands almost radially (Fig. \ref{Fig2}(b))
and rushes toward the gas disc (Fig. \ref{Fig2}(c)).
At $\sim 0.47 \mbox{ years}$, as seen in Fig. \ref{Fig2}(d),
the bulk of flow is bent to go through the disc. 
In this stage, some of gas disc is stripped out, but
much of the blast wave energy can escape owing to the bending 
of gas flow. On the other hand, 
the flow directed to the disc center is strongly
decelerated by the collision with the disc, and begins to
deform the disc (Fig. \ref{Fig2}(e)(f)).

Figure \ref{Fig3} shows the deformation and stripping of the disc 
at later stages. The figure zooms in the disc regions. 
The rotation period at the outer edge of gas disc is 
$\sim 148 \mbox{ years}$. 
Fig. \ref{Fig3} (a) is the initial state again. 
As seen in Figures \ref{Fig3} (b) and (c),
the blast wave pushes the left-side edge of gas disc and produces a dent
within a few years, because the time-scale of the SN explosion is 
much shorter than the rotation period of the outer edge of gas disc. 
The rotation of disc transfers this dent, resulting in the overall deformation 
of the disc (Fig. \ref{Fig3} (d)). 
Through this deformation, the rotation balance breaks down,
and resultantly some of gas is blow out by centrifugal force,
as seen in Figure \ref{Fig3} (e).
Eventually, the gas disc is settled in a quasi-steady state 
within about $100 \mbox{ years}$ (Fig. \ref{Fig3} (f)).
As a result, roughly 70\% of an original disc around the BH is left over.
In Figure \ref{Fig4}, the deformation and stripping processes
are shown with three-dimensional volume rendering visualization.
The survival of the gas disc after the heaving and ruffling 
by the blast wave collision is clearly shown. 

Previously, \citet{Wheeler+75} argued that in a binary system 
the mass stripping and ablation from a star by the impinging balastwave 
is well expressed by a non-dimensional parameter $\Psi$, which is defined as 
\begin{equation}
\Psi \equiv \frac{1}{4}\frac{M_{\rm SN}}{M_{\rm c}}
\frac{R^{2}}{a^{2}}\left(
\frac{v_{\rm SN}}{v_{\rm es}}-1
\right), 
\end{equation}
where $M_{\rm SN}$ is the mass of gas expelled by a SN, 
$M_{\rm c}$ and $R$ are respectively the mass and radius 
of object affected by blast wave, 
$v_{\rm SN}$ is the typical speed of the blast wave,  
$v_{\rm es}$ is the escape velocity. 
They obtained the mass fraction ejected by the stripping and ablation
as a function of $\Psi$. 
We can evaluate $\Psi$ in the present system, where
$M_{\rm SN}=50 M_{\odot}$, $M_{\rm c}=M_{\rm disc}=30 M_{\odot}$, 
$R=280 \mbox{AU}$, $a=700 \mbox{AU}$ 
and $v_{\rm SN}=4000 \mbox{ km}/\mbox{sec}$. 
$v_{\rm es}$ is approximated to be
\begin{equation}
v_{\rm es}=\left(\frac{2 G M_{\rm BH}}{R}\right)^{1/2},
\end{equation}
where $M_{\rm BH}=1000 M_{\odot}$. 
Then, we find $\Psi \simeq 3.28$ in our model A1. 
According to the criterion for a polytropic star with $n=3$ given
by \citet{Wheeler+75}, it is predicted that about a half of mass is 
ejected in the case of $\Psi \simeq 3$. 
In the present simulation, about 30\% of mass is ejected eventually.
Therefore, we can conclude that $\Psi$ is a fairly good measure for
the mass ejection even in a disk system. However, in the disk system,
a little more mass remains than the prediction for a spherical star.
In the collision of blast wave with a gas disk, the streamline is
more strongly bent above and below the disc. Hence,
the momentum transferred to the gas in a disk is reduced, and
more mass can be left over. 
Therefore, the difference from the spherical prediction
can be reasonably understood in terms of the geometrical effect.

\subsection{Dependence on Binary Separation}

In order to investigate the dependence on the binary separation,
we perform additional two simulations with $a=350 \mbox{ AU}$ (model B1),
which is a half separation of model A1,
and with $a=140 \mbox{ AU}$ (model B2), which is a one-fifth 
separation of model A1. 
We assume the mass of the disc to be the same as model A1. 
Therefore, the disc radius becomes smaller
and the density is higher for these models. 
Also, the energy of the SN explosion, 
the temperature, and the total mass of ejected gas 
to be the same as model A1.
The rotation period of the outer edge of gas disc for model B1 and B2 is 
$\sim 52 \mbox{ years}$ and $\sim 13 \mbox{ years}$, respectively. 
Figure \ref{Fig5} shows the density distributions and the velocity fields
in the final quasi-steady state, which is reached 
at $t=38.19 \mbox{ years}$ and $11.95 \mbox{ years}$ 
for models B1 and B2, respectively. 
Hence, the time to reach the final state is shorter 
for the shorter separation. 
Interestingly, it is found that more mass is left over for
shorter binary separation. Roughly 80 \% of disc survives for
model B1, and nearly 85 \% of disc does for model B2.
This can be understood in terms of the binding energy $E_b$
of the disc relative to the kinetic energy $E_k$ of 
the SN blast wave. The ratio $E_b/E_k$ is $\sim 0.1$ and
$\sim 0.2$ for models B1 and B2, respectively, while
$E_b/E_k \sim 0.03$ for model A. Thus, $E_b/E_k$ is
thought to be a key physical quantity that determines
the remaining mass of disc. This point is argued again later.

\subsection{Dependence on Density Distribution}

To check whether the deformation and stripping of disc depends on 
the density distribution of gas disc, simulations for discs 
with different density distributions are performed.
One is a flat density distribution model with $n=0$ on the equatorial plane 
(model C1), and the other is a centrally concentrated model with
$n=-2$ (model C2). 
The mass of gas disc is the same as model A1. 
Figure \ref{Fig6} shows the results for a flat density distribution disc
 (model C1).
The snapshots of density distributions and velocity fields 
at $t=0.00, 10.32, \mbox{ and } 110.16 \mbox{ years}$ are presented. 
The results look similar those shown in Figure \ref{Fig3} at 
the same evolutionary stages. Actually, the remaining mass is not
changed significantly. 
Figure \ref{Fig7} show the results for a centrally-concentrated disc (model C2).
The snapshots of density distributions and velocity fields 
at $t=0.00, 10.37, \mbox{ and } 105.69 \mbox{ years}$ 
are presented. 
It is found that, for model C2, the influence by the blast wave is
slightly weaker in the central regions of the disc (Fig. \ref{Fig7} (b)), 
compared to model C1 (Fig. \ref{Fig6} (b)). 
But, the remaining mass increases by only several \%. 
Therefore, we can conclude that the density distributions of disc
do not influence the final mass of disc strongly.

\subsection{Remaining Mass}

In Figure \ref{Fig8}, the time sequences of the disc mass 
are summarized by the ratio of the remaining disc mass 
to the disc mass without a supernova explosion of companion star. 
For model A1, C1 and C2, the disc is settled in quasi-steady states 
in $60$ years,
and $\sim 70 \%$ of disc gas remains, almost regardless of density 
distributions of disc. 
For model B1 ($a=350 {\rm AU}$), a quasi-steady state is attained 
in $20$ years and
$\sim 80 \%$ of disc gas remains.
For model B2 ($a=140 {\rm AU}$), the disc reaches 
a quasi-steady state in $10$ years 
and $\sim 85 \%$ of disc gas remains.

As argued in Section \ref{deformation}, the remaining amount of gas 
can be understood in terms of the ratio of $E_{b}$ to $E_{k}$, 
where $E_{b}$ is the binding energy of the gas disc around the BH and  
$E_{k}$ is the kinetic energy of the blast wave 
taking into account the solid angle of the disc viewed 
from the SN explosion point. 
$E_{b}$ is evaluated by integrating potential energy 
of the gas around the BH.
Figure \ref{Fig9} shows the remaining mass fraction against
the $E_{b}/E_{k}$ ratio. 
In this diagram, two more simulations are added with changing 
the disc mass: model D1 has the disc mass of $10 M_{\odot}$, 
which is one-third of that in model A, and model D2 has 
the disc mass of $3 M_{\odot}$, which is one-tenth of that in model A.
Obviously, the remaining mass fraction correlates positively with 
$E_{b}/E_{k}$. 
Thus, a key physical parameter that determines the remaining
mass fraction is the $E_{b}/E_{k}$ ratio.
Interestingly, even if the binding energy is less than 1\% of the 
kinetic energy of SN blast wave, a considerable amount of mass
remains. This is a three-dimensional effect of deformation and
stripping, which is demonstrated in Figure \ref{Fig4}.

\section{Conclusions}
\label{Sec4}

We have explored the dynamical impact of SN explosion in a very massive binary
system. We have simulated the deformation and stripping of a gas disc
around a BH, using three-dimensional generalized curvilinear coordinates.
Also, the dependence on the binary separation, the density distribution of disc,
and the mass of disc have been studied. 
As a result, we have found that a SN blast wave is not so devastating that
a disc around a BH is totally evaporated. 
It has turned out that the remaining mass fraction is basically determined
by the ratio of the disc binding energy to the SN kinetic energy. 
As a matter of importance, even if the binding energy of gas disc 
is much smaller than the kinetic energy of blast wave, 
a significant amount of mass can remain. For instance,
when the binding energy is about 1\% of the 
kinetic energy of SN blast wave, about a half of mass is left over.
We have found that the three dimensionality of the disc deformation is 
essential to make a disc around a BH survive. 

The present results imply that in a massive binary system like
a Pop III binary, the BH accretion activity can be long-lived 
even after a massive companion star explodes. 
Such BH accretion can emit ultraviolet radiation, and therefore
a Pop III BH accretion disc can be an important candidate 
as a reionization source in the early universe.

\section*{Acknowledgments}
We thank K. Hiroi for many useful comments. 
Numerical simulations have been performed with computational facilities 
at Center for Computational Sciences in University of Tsukuba. 
This work was supported in part by Grants-in-Aid, Specially
Promoted Research, 16002003, from MEXT in Japan.



\setcounter{figure}{0}
\clearpage
\begin{table*}
\caption[]{Adopted parameters for model calculations. 
In this table, $a$ represents the binary separation, 
$M_{\rm disc}$ does the mass of gas disc, 
and $n$ does the power-law index of density distributions 
on the equatorial plane. 
}
\centering
\begin{tabular}{r|rrrl}
\hline
Model & $a$ [AU] & $M_{\rm disc}$ & $n$ & Remarks \\
\hline
A1 & $700$ & $30$ & $-1$ & Fiducial model \\
B1 & $350$ & $30$ & $-1$ & Half of the separation in model A1 \\
B2 & $140$ & $30$ & $-1$ & One-fifth of the separation in model A1 \\
C1 & $700$ & $30$ & $0$ & Flat density distribution disc \\
C2 & $700$ & $30$ & $-2$ & Centrally concentrated density distribution disc \\
D1 & $700$ & $10$ & $-1$ & One-third of the disc mass in model A1 \\
D2 & $700$ & $3$ & $-1$ & One-tenth of the disc mass in model A1 \\
\hline
\end{tabular}
\label{tab1}
\end{table*}

\begin{figure*}
\centerline{\psfig{figure=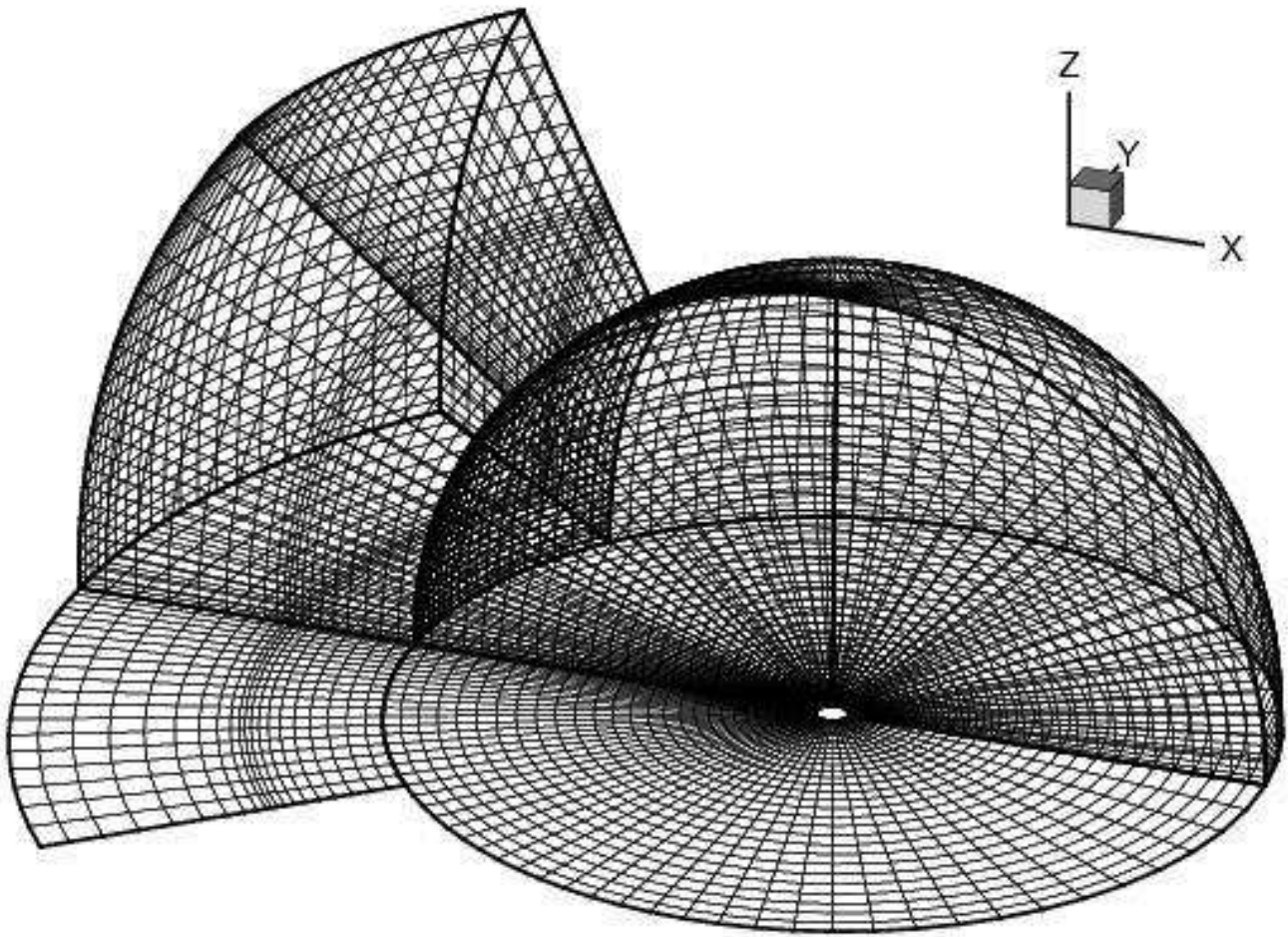,clip=ture,width=\textwidth}}
\caption[dummy]{
The three-dimensional generalized curvilinear coordinate system 
used in this paper. 
The bottom plane is the equatorial plane. 
The center of the gas disc, i.e. the BH, is placed at the center 
of a rightward hemisphere, and the SN explosion occurs at the center
of an adjacent protuberant region. A half of grids in the computational
domain are shown, after the whole domain is cut by a plane 
including the BH and the SN explosion point perpendicular 
to the equatorial plane. 
The number of grid points is $51 \times 101 \times 31$.
}
\label{Fig1}
\end{figure*}

\begin{figure*}
\centerline{\psfig{figure=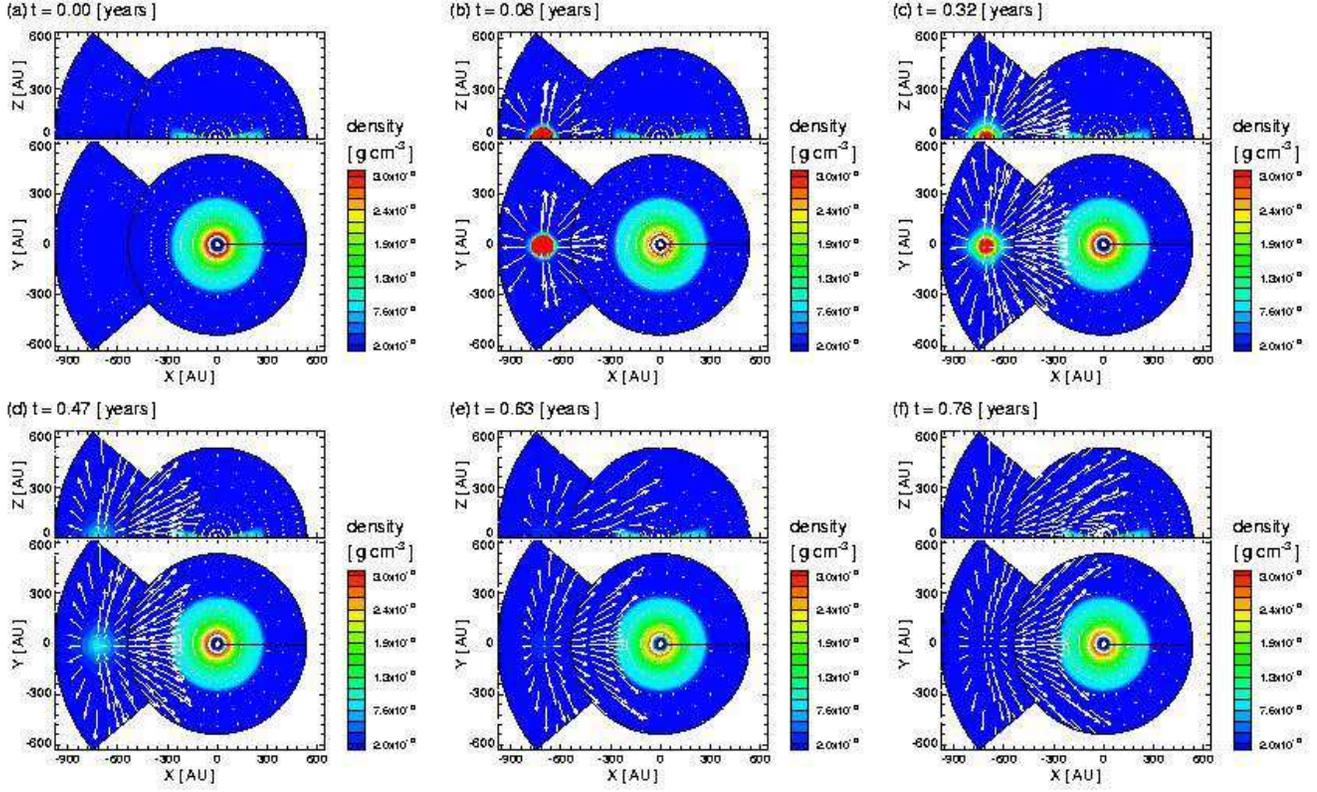,clip=ture,width=\textwidth}}
\caption[dummy]{
Early stages of the interaction of a SN blast wave with a disc around a BH 
for model A1. 
Snapshots show density distributions and velocity fields in the $x-y$ plane 
with $z=0$ ({\it lower panels}), 
and in $x-z$ plane with $y=0$ ({\it upper panels}).
}\label{Fig2}
\end{figure*}

\begin{figure*}
\centerline{\psfig{figure=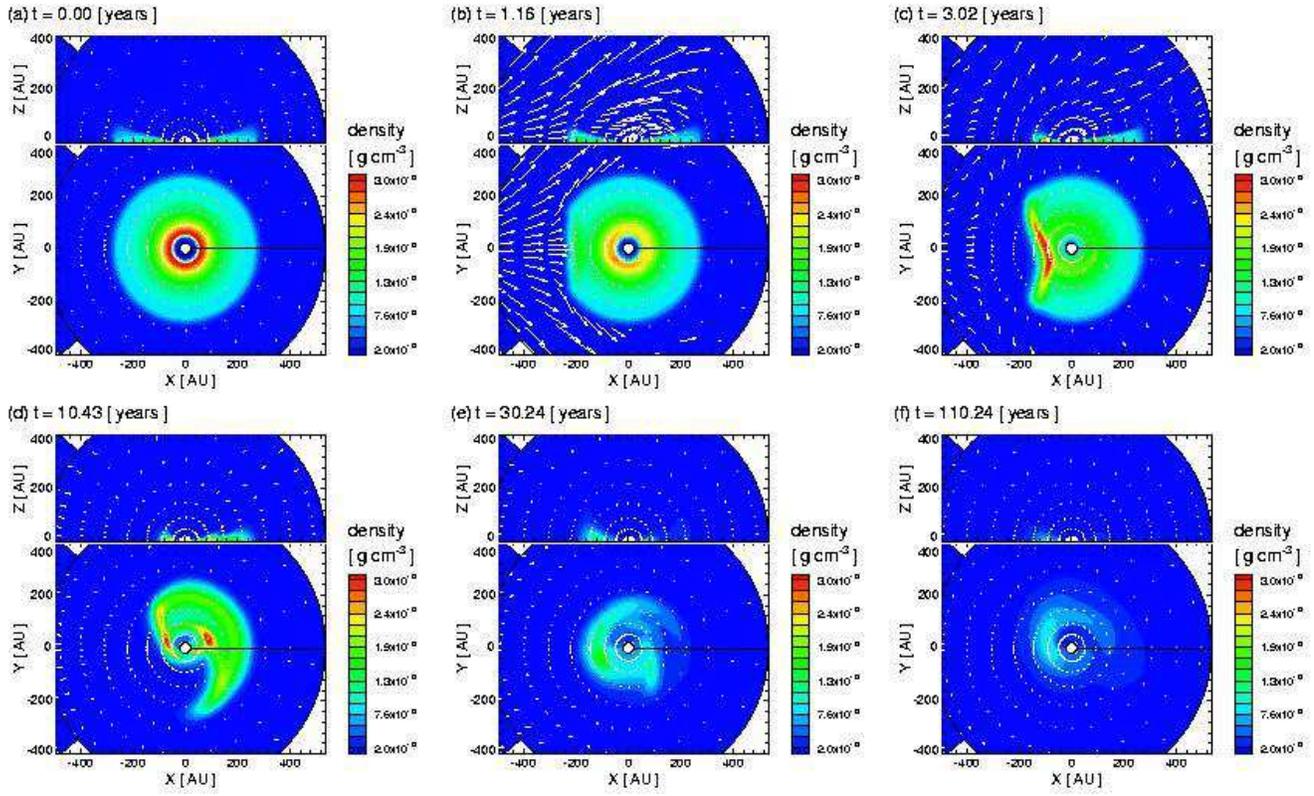,clip=ture,width=\textwidth}}
\caption[dummy]{
Later stages of the interaction of a SN blast wave with a disc around a BH 
for model A1, where
the deformation and stripping processes of the gas disc are shown.
The density distributions and the velocity fields are shown with
zooming in disc regions. 
}\label{Fig3}
\end{figure*}

\begin{figure*}
\centerline{\psfig{figure=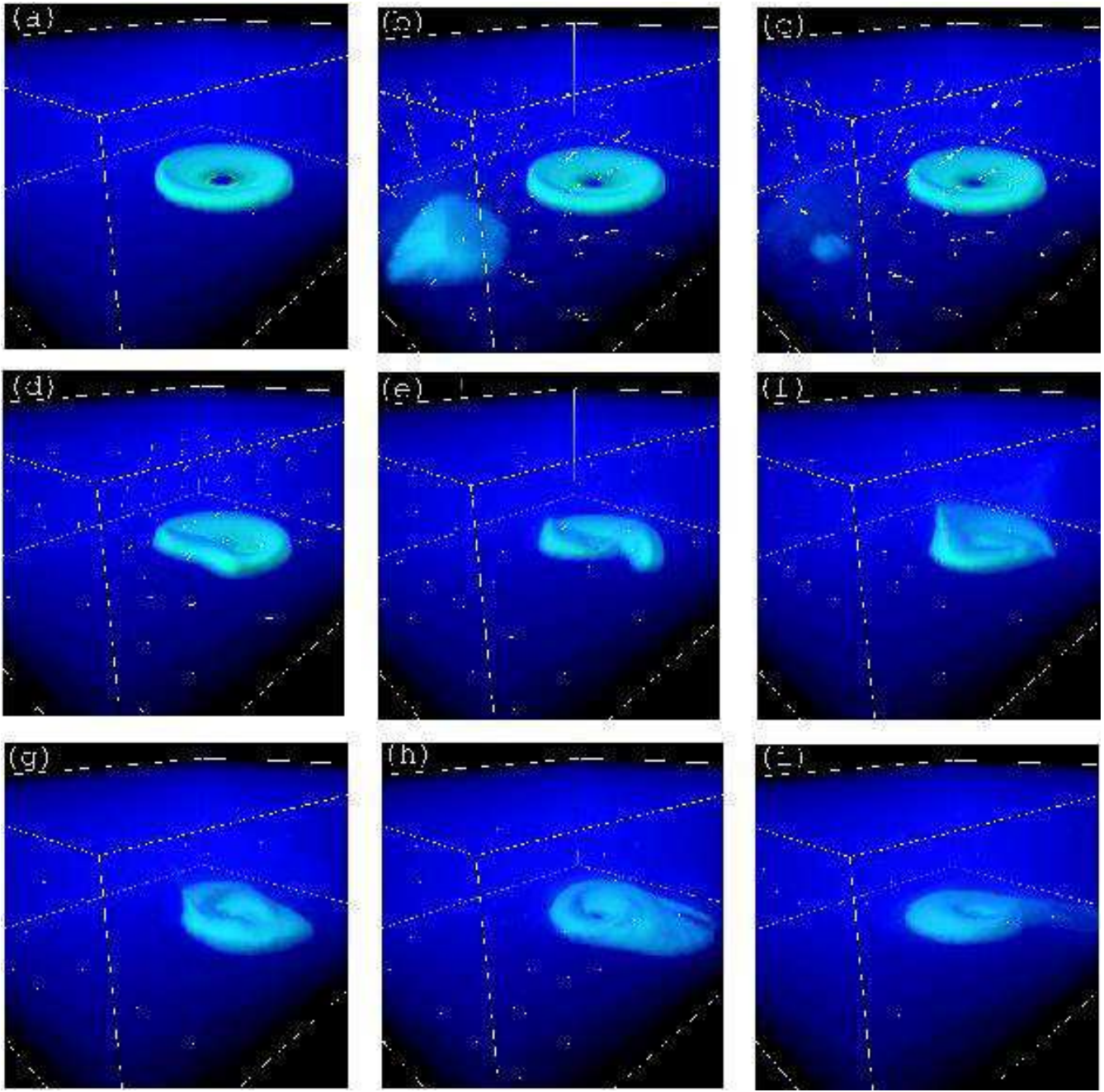,clip=ture,width=\textwidth}}
\caption[dummy]{
3D volume rendering visualization of density distributions 
with velocity fields for model A1. 
Figures (a), (b), (c), (d), (e), (f), (g), (h) and (i) 
represent the snapshots at $t=0.00, 0.63, 0.78, 2.90, 11.17, 
35.72, 53.98, 72.98, \mbox{ and } 110.24$ yrs, respectively.
}\label{Fig4}
\end{figure*}

\begin{figure*}
\centerline{\psfig{figure=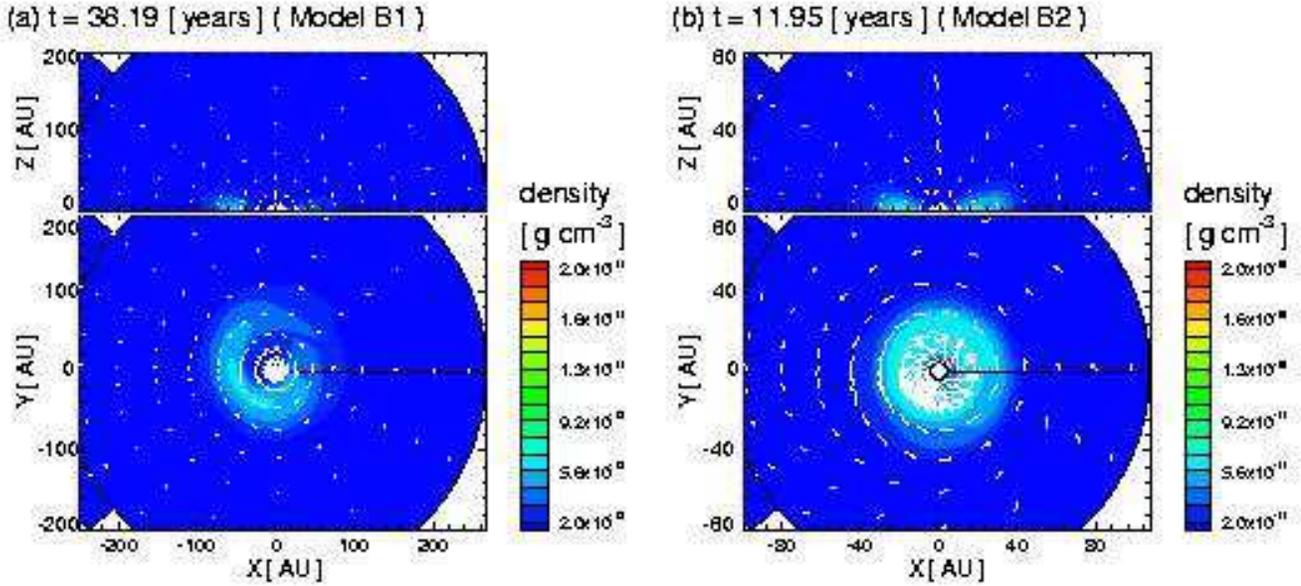,clip=ture,width=\textwidth}}
\caption[dummy]{
The density distributions and the velocity fields 
in the final state for shorter separation models; model B1 is a model
with $a=350 \mbox{ AU}$ ({\it left panel}) and model B2 is a model
with $a=140 \mbox{ AU}$ ({\it right panel}).  
The snapshots at $t=38.19 \mbox{ years}$ and $11.95 \mbox{ years}$ 
are presented for model B1 and B2, respectively. 
}\label{Fig5}
\end{figure*}

\begin{figure*}
\centerline{\psfig{figure=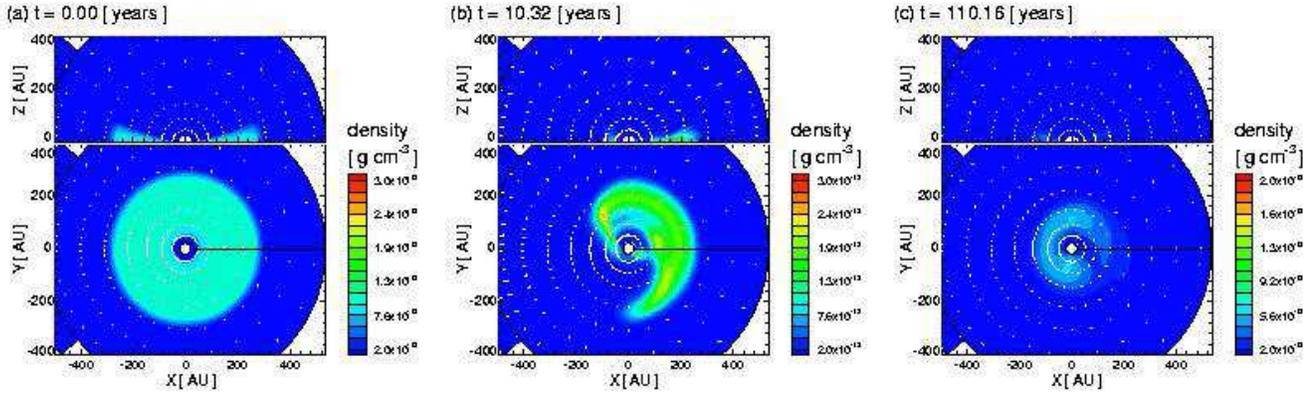,clip=ture,width=\textwidth}}
\caption[dummy]{
Same as Fig. \ref{Fig3} but for a flat density distribution model 
with $n=0$ (model C1). 
The snapshots at $t=0.00, 10.32, \mbox{ and } 110.16$ years are presented. 
}\label{Fig6}
\end{figure*}

\begin{figure*}
\centerline{\psfig{figure=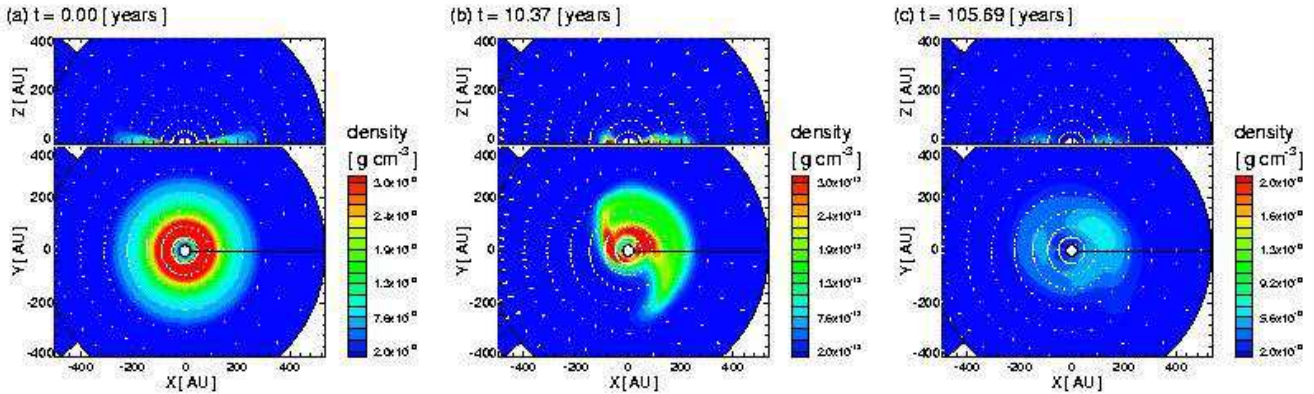,clip=ture,width=\textwidth}}
\caption[dummy]{
Same as Fig. \ref{Fig3} but for a central-concentration 
density distribution model with $n=-2$ (model C2).
The snapshots at $t=0.00, 10.37, \mbox{ and } 105.69$ years are presented. 
}\label{Fig7}
\end{figure*}

\begin{figure*}
\centerline{\psfig{figure=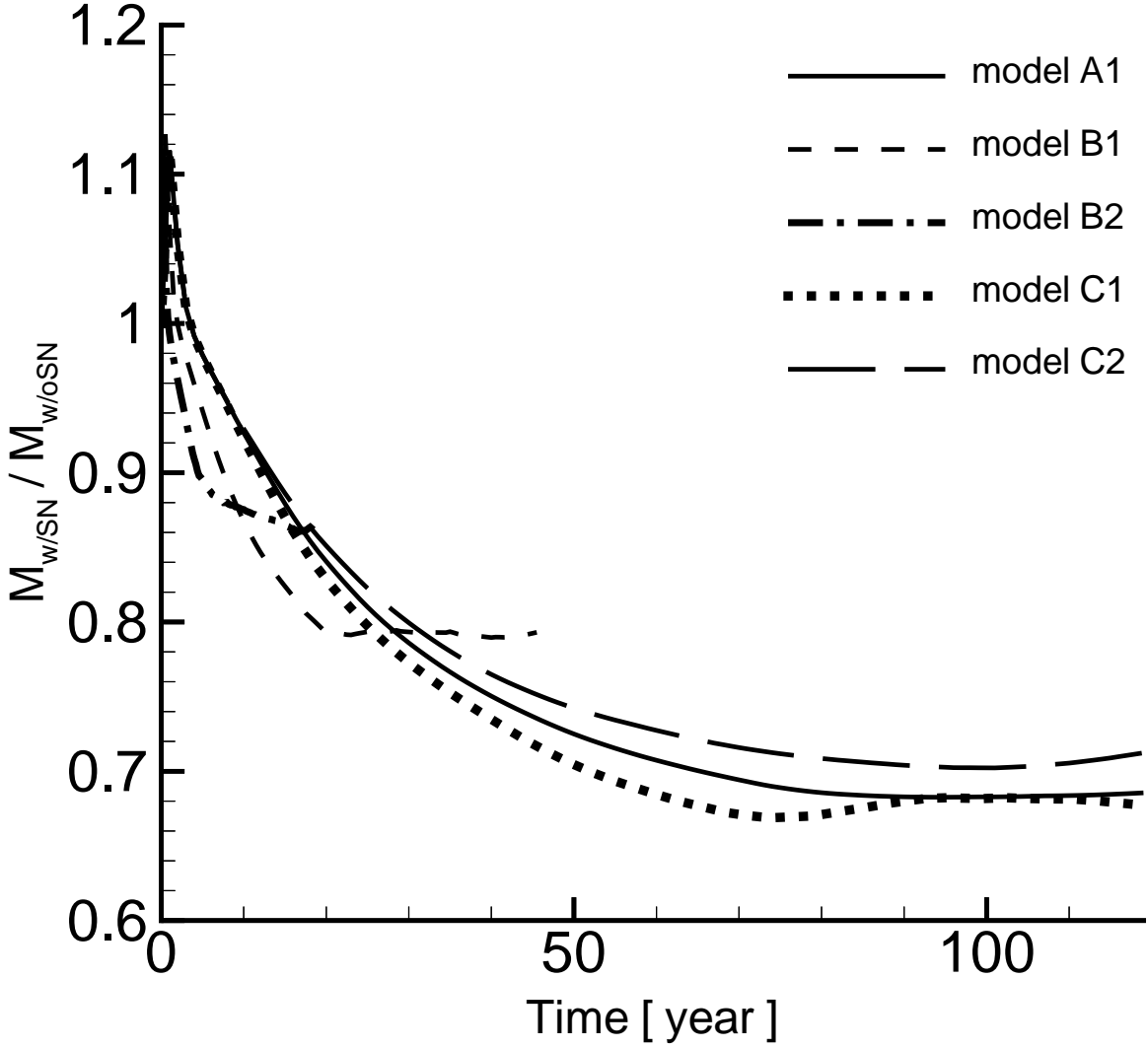,clip=ture,width=\textwidth}}
\caption[dummy]{
The time variation of the remaining mass of a disc affected 
by SN explosion, where the ratio of remaining mass to the mass 
of an undisturbed disc is shown. 
The horizontal axis is time in units of years. 
A solid line, a dashed line, a dash-dotted line, a dotted line and 
a log-dashed line show model A1, B1, B2, C1 and C2, respectively. 
Note that the ratio for each case starts exactly from unity.
But, the strong inflow by supernova explosion causes the transient
rapid growth of ratio by about 10\% at the very early stage.
}
\label{Fig8}
\end{figure*}

\begin{figure*}
\centerline{\psfig{figure=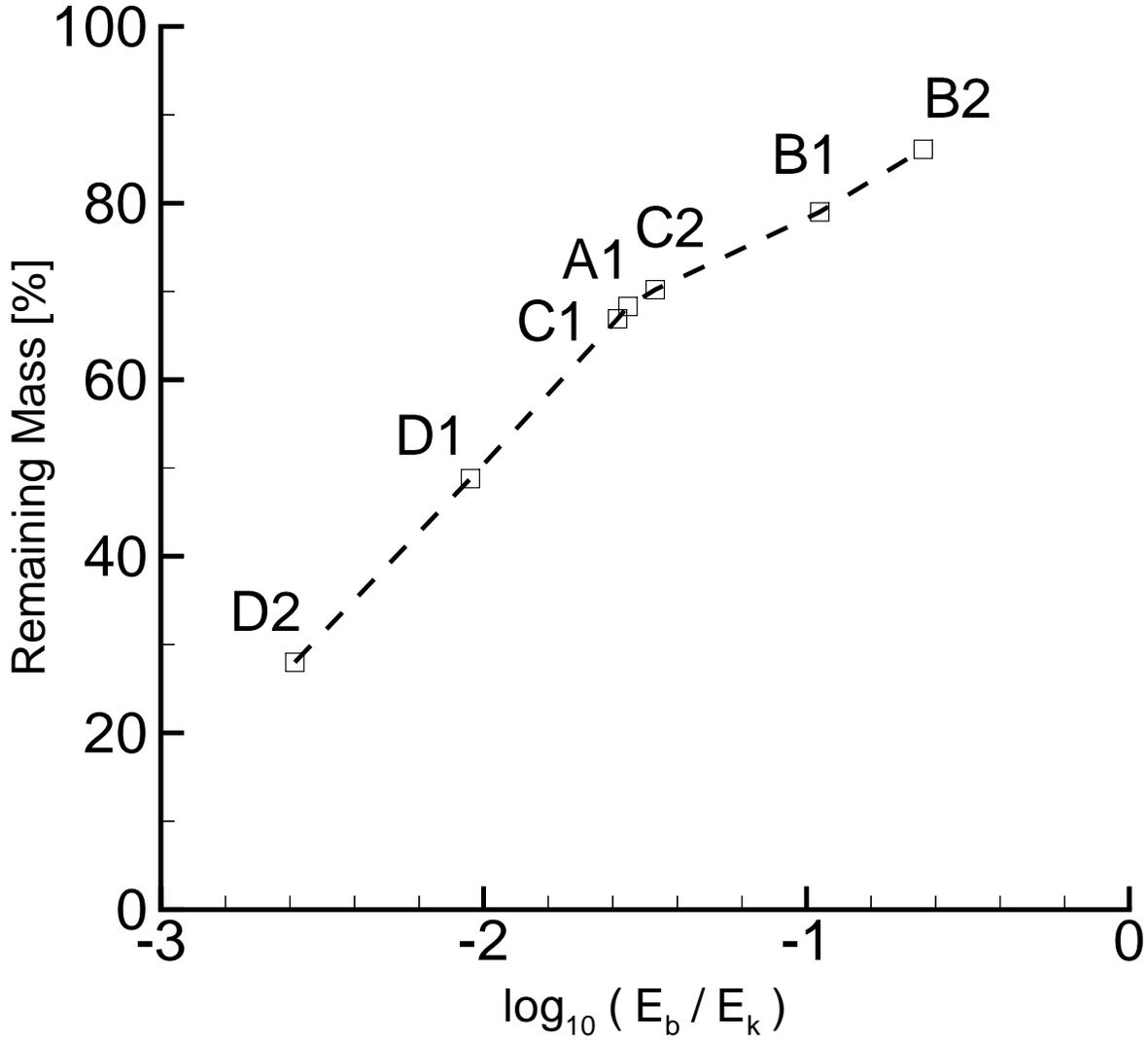,clip=ture,width=\textwidth}}
\caption[dummy]{
The resultant remaining mass fraction is shown against 
the ratio of $E_{b}$ to $E_{k}$, 
where $E_{b}$ is the binding energy of the gas disc around a BH and  
$E_{k}$ is the kinetic energy of blast wave 
taking into account the solid angle of the disc viewed 
from the SN explosion point. 
The names near the symbols are model names.  
}
\label{Fig9}
\end{figure*}

\end{document}